\documentclass{article}

\usepackage{caption}
\usepackage{PRIMEarxiv}
\usepackage{graphicx}
\usepackage{amsmath}
\usepackage{amsfonts} 
\usepackage{times}  % DO NOT CHANGE THIS
\usepackage{makecell}
\usepackage{helvet}  % DO NOT CHANGE THIS
\usepackage{courier}  % DO NOT CHANGE THIS
\usepackage[hyphens]{url}  % DO NOT CHANGE THIS
\usepackage{graphicx} % DO NOT CHANGE THIS
\usepackage[ruled,vlined]{algorithm2e}
\usepackage{algorithmic}
\usepackage{xcolor}
\usepackage{bm}
\usepackage[utf8]{inputenc} % allow utf-8 input
\usepackage[T1]{fontenc}    % use 8-bit T1 fonts
\usepackage{hyperref}       % hyperlinks
\usepackage{url}            % simple URL typesetting
\usepackage{booktabs}       % professional-quality tables
\usepackage{amsfonts}       % blackboard math symbols
\usepackage{nicefrac}       % compact symbols for 1/2, etc.
\usepackage{microtype}      % microtypography
\usepackage{lipsum}
\usepackage{fancyhdr}       % header
\usepackage{graphicx}       % graphics
\graphicspath{{media/}}     % organize your images and other figures under media/ folder
\newcommand{\RN}[1]{\uppercase\expandafter{\romannumeral#1\relax}}
\usepackage{natbib} 
\newcommand{\Qtot}{Q^{\phi}_{\text{tot}}}
\newcommand{\Qtotp}{Q^{\phi'}_{\text{tot}}}
\newcommand{\kE}{\kappa_{e,\mathcal{Q}_{i}}^{\phi_i}}
\newcommand{\E}{\mathbb{E}}
\newtheorem{lemma}{Lemma}

\title{Ensemble-MIX: Enhancing Sample-Efficiency in Multi-Agent RL
via Selective Ensemble-Based Exploration}

\author{
  Tom Danino \\
  The Andrew and Erna Viterbi Faculty of Electrical \& Computer Engineering \\
  Technion- Israel Institute of Technology \\
  \texttt{tdanino12@gmail.com} \\
   \And
  Nahum Shimkin \\
  The Andrew and Erna Viterbi Faculty of Electrical \& Computer Engineering \\
  Technion- Israel Institute of Technology\\
  \texttt{shimkin@ee.technion.ac.il} \\
}
\begin{document}
\maketitle
\begin{abstract}
Multi-agent reinforcement learning (MARL) has achieved state-of-the-art performance across a variety of cooperative tasks. However, MARL algorithms typically require substantially more environment interactions than their single-agent counterparts to converge, largely attributed to the challenges of exploring a large joint action space inherent to multi-agent environments. To address these issues and improve sample efficiency, we propose extending multi-agent policy gradient (MAPG) methods with an ensemble of critics and leveraging it for selective, noise-aware exploration. Our proposed technique first derives from the ensemble an exploration signal that identifies high-uncertainty states and actions. The signal is then incorporated into the behavior policy via a novel actor-logit manipulation algorithm that encourages exploration toward uncertain regions. We further identify a key obstacle in adapting ensemble-based exploration to MARL: joint learning of multiple agents increases stochasticity and amplifies aleatoric noise within the ensemble, resulting in excessive or misdirected exploration. We analyze this effect on standard uncertainty measures, including ensemble variance and voting mechanisms, and introduce a novel kurtosis-based signal that selectively emphasizes areas of higher tail disagreement within the ensemble, while in many cases remaining less sensitive to noise-induced over-exploration, as our analysis shows.
The evaluation demonstrates that our method outperforms state-of-the-art baselines on standard MARL benchmarks, including a large collection of SMAC maps.

\end{abstract}
\keywords{Multi-agent, Reinforcement learning, Ensemble learning, Selective exploration}

\section{Introduction}
In recent years, \emph{centralized training with decentralized execution} (CTDE) \cite{rashid2020weighted} has achieved state-of-the-art performance across challenging multi-agent benchmarks \cite{qin2025gradps, chen2025novelty,li2025learning}. CTDE enables parameter and information sharing during training while maintaining decentralized execution at test time. A key factor of its success is driven by value decomposition, where a centralized mixing network decomposes a global value function into per-agent components, facilitating coordinated learning. This idea has also been extended to decompose the critic network in multi-agent policy gradient (MAPG) methods \cite{wang2020dop, zhou2022pac}, yielding significant performance gains.

Despite this progress, CTDE methods remain susceptible to the curse of dimensionality \cite{du2021survey}, as the joint action space expands exponentially with the number of agents, leading to longer training times. Consequently, standard exploration techniques from single-agent RL such as entropy maximization scale poorly, often leading to wasteful and redundant exploration \cite{kim2023adaptive}. This motivates the examination of alternative exploration paradigm: uncertainty-based exploration, particularly ensemble driven, which fits this problem more naturally due to the ensembles ability to capture epistemic uncertainty (lack of knowledge) and underexplored areas \cite{lee2021sunrise}.

Leveraging these capabilities, we utilize the ensemble technique to tackle the problem of efficient exploration in MARL, which remains a key bottleneck for achieving good sample efficiency in such high-dimensional settings. While ensembles have proven effective in single-agent RL, our study reveals that adapting them to MARL presents a critical yet overlooked challenge: High aleatoric noise arising from suboptimal and exploratory behaviors of multiple learning agents \cite{wang2020dop,kuba2021settling} shifts uncertainty estimates, leading to excessive exploration of noisy states. 
Although prior work has identified noise-induced over-exploration as a (manageable) limitation of ensembles in single-agent RL \cite{liu2024ovd}, we demonstrate that it is significantly amplified in MARL due to interactions between multiple learning agents, with potentially severe implications for exploration.

%To better understand this phenomenon, we first examine a simple motivating example. On an $8\times8$ grid, taxis deliver passengers via two-leg handoffs along routes that cross a central road (in red), which inflicts a random penalty whenever two or more taxis are on it at once. This penalty is aleatoric, it depends on unpredictable taxi co-occupations and other agent exploring the terrain. As we increase the number of taxis and passengers (4, 6, 8 agents), the read area increasingly dominates ensemble L2 disagreement on the grid, while epistemic disagreement elsewhere almost doesnt change. In the motivation section, we show this problem persists behind the L2 disagreement, and can be observed in common measurements used for exploration such as ensemble variance and voting mechanisms. We leverage this insight in selecting an exploration signal that is both efficient and robust to noise-induce settings. 

\begin{figure*}[t]
\centering
\includegraphics[width=0.6\columnwidth]{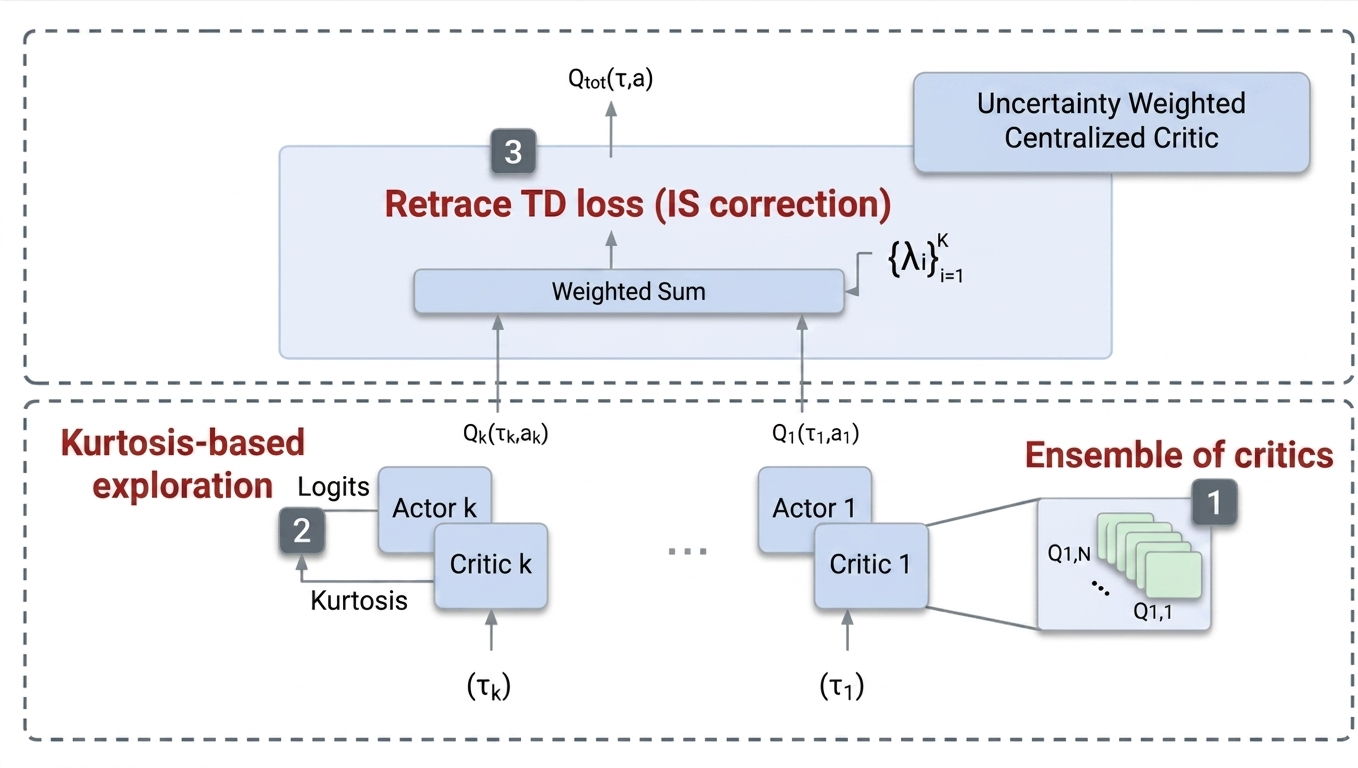}
\caption{Ensemble MIX architecture: (1) Each agent's critic is an ensemble of $N$ sub-critics. (2) The
ensemble kurtosis utilized for selective exploration during action selection. (3) The backbone follows DOP with retrace instead of TB.}
\label{fig:architecture}
\end{figure*}

In this work, we present \emph{Ensemble MIX}, an ensemble-based MAPG algorithm designed for selective exploration of large joint action spaces. Ensemble MIX consists of two main components: (1) A decentralized ensemble of critics integrated into the mixing network (Fig. \ref{fig:architecture}, component 1), utilized to compute a selective exploration signal. (2) An exploration mechanism that uses this signal to manipulate the actor’s logits toward actions with high uncertainty (component 2). To construct the signal from ensemble disagreement, we introduce a novel approach based on ensemble kurtosis. The kurtosis signal improves exploration efficiency by prioritizing regions with high disagreement and heavier distribution tails. Moreover, our analysis shows that it is less prone to noise-induced over-exploration, alleviating the challenge of amplified noise in MARL.

%\begin{figure}[t]
%\centering
%\includegraphics[width=0.4\textwidth]{aaai2027/images/motivational_fig.png}
%\caption{\textbf{Multi-taxi environment.} (a) Environment illustration. (b) As the number of agents increases, ensemble disagreement on noisy areas (red) become more dominant compared to destination (green) and regular cells (blue).}
%\label{fig:toy0}
%\end{figure}

Ensemble MIX is simple to implement and incorporate into an existing MAPG framework; specifically, we build our design upon DOP \cite{wang2020dop} as the underlying actor–critic backbone. Note that, to mitigate the distributional shift introduced by incorporating kurtosis into the behavioral policy, we replace the TB loss used in DOP for the critic learning with Retrace (Fig. \ref{fig:architecture}, component 3), enabling importance-sampling (IS) correction for off-policy data. Specifications of retrace and DOP are given in the background section. 

Ensemble-MIX differs from prior work on exploration and ensemble learning in two key aspects. First, unlike methods such as SUNRISE, EMAX, and TEEN \cite{lee2021sunrise, learningensemble, li2023keep}, which inject exploration through selection over uncertainty-augmented Q-values or ensemble diversity, our approach operates directly on the actor's logits, applying a smooth shift to the behavior policy that integrates naturally with the policy-gradient setting, thereby incorporating ensemble-based exploration into MAPG. Second, to the best of our knowledge, we are the first to identify the problem of noisy ensemble predictions in MARL and mitigate it via a designated uncertainty signal. A detailed comparison is provided in the related work.

Our work makes the following main contributions:
\begin{itemize}
\item We introduce Ensemble MIX, a novel MAPG algorithm for sample-efficient exploration through manipulation of actor logit based on an exploration signal. 
\item To derive the exploration signal, we propose augmenting MAPG with a decentralized critic ensemble.
\item We identify and study a key challenge in MARL ensembles: multi-agent stochasticity amplifies disagreement in areas of aleatoric noise, degrading exploration.
\item To tackle the noisy action-state space, we introduce a novel kurtosis-based signal and provide an analysis of its ability to distinguish epistemic and aleatoric uncertainty.
\end{itemize}
The evaluation section demonstrates the effectiveness of Ensemble MIX on the StarCraft \cite{samvelyan2019starcraft}, Predator-Prey \cite{zhou2022pac}, and the multi-agent car-following \cite{shen2023riskq} environments.

\section{Background \label{sec:back}}

\subsubsection{Dec-POMDPs.}
A decentralized partially observable MDP models cooperative agents as a tuple
$\langle\mathcal{I},\mathcal{S},\mathcal{A},\mathcal{P},\Omega,\mathcal{O},r,\gamma\rangle$,
where $\mathcal{I}$ is a set of $k$ agents and $s\in\mathcal{S}$ is the true state. Each
agent selects $a_i\in\mathcal{A}$, forming the joint action $\bm{a}=[a_i]_{i=1}^{k}$,
with shared reward $r(s,\bm{a})$ and transitions $\mathcal{P}(s'|s,\bm{a})$. Agents
receive observations $o_i\in\Omega$ through $\mathcal{O}(o_i|s,a_i)$ and discount future
rewards by $\gamma\in[0,1)$. Joint action--observation histories are
$\bm{\tau}\in\mathcal{T}$, with policies $\{\pi_{\theta_i}\}_{i=1}^{k}$, behavior
policies $\{\mu_i\}_{i=1}^{k}$, and actor logits $\bar z^i=f_{\theta_i}(\tau_i)$.

\textbf{Value decomposition and monotonic constraints in multi-agent reinforcement learning.}
Value decomposition methods decompose the joint action-value function into per-agent value functions while satisfying the Individual-Global-Maximum (IGM) condition:
$
\arg\max_{\mathbf{a}} Q_{\mathrm{tot}}(\boldsymbol{\tau}, \mathbf{a})
=
\bigl(\arg\max_{a_i} Q_i(\tau_i,a_i)\bigr)_{i=1}^{k}.
$ VDN achieves this by summing individual $Q$-values, whereas QMIX uses a monotonic mixing network to model a more expressive non-linear joint value function while preserving IGM.

\subsubsection{DOP backbone.}
We adopt DOP~\citep{wang2020dop} as our backbone. DOP is an MAPG method that combines a centralized decomposed
critic with decentralized actors. They train the critic with a combination of on-policy
and off-policy losses,
\begin{equation}
\mathcal{L}_{\text{critics}}(\phi)=\mathbf{c}\,\mathcal{L}^{\text{on-TD}(\lambda)}(\phi)
+(1-\mathbf{c})\,\mathcal{L}^{\text{off-TB}}(\phi),
\label{eq:dop_loss}
\end{equation}
with the off-policy tree-backup target
\begin{align}
y^{\text{off}}=&\,\Qtotp(\bm{\tau},\bm{a})
+\sum_{t=0}^{m-1}\prod_{l=0}^{t}\Big(\lambda\bm{\pi}(\bm{a}_l|\bm{\tau}_l)\Big)
\big[r_t \nonumber\\
&+\gamma\E_{\bm{a}_{t+1}\sim\bm{\pi}}[\Qtotp(\bm{\tau}_{t+1},\bm{a}_{t+1})]
-\Qtotp(\bm{\tau}_t,\bm{a}_t)\big].
\label{eq:dop_off}
\end{align}
$\mathcal{L}^{\text{off-TB}}(\phi)$ is defined as $\mathrm{E}_{\bm{\pi}}[(y^{\text{off}}-Q_{\text{tot}}^{\phi}(\bm{\tau},\bm{a}))^2]$. The product $\prod_{l=0}^{t} \lambda\pi(\bm{a}_l|\bm{\tau}_l)$ down-weights a transition whenever the target policy is unlikely to have chosen the observed actions, reducing the influence of off-policy data. The full specification of the critic loss, along with details on other components, can be found in \cite{wang2020dop}.

\subsubsection{Retrace.}
Retrace($\lambda$) \cite{munos2016safe} replaces the TB coefficient $\lambda\pi(a_s|\tau_s)$ with the truncated importance ratio $c_s = \lambda \min\!\left(1, \frac{\pi(a_s|\tau_s)}{\mu(a_s|\tau_s)}\right)$. Truncating at $1$ caps the variance that plain importance sampling suffers from, while leaving traces uncut in the near-on-policy regime so that full returns are still used. For any coefficient $0 \le c_s \le \frac{\pi}{\mu}$, the operator is a $\gamma$-contraction with fixed point $Q^\pi$ for arbitrary behavior policies $\mu$, as long as $\mu$ assigns positive probability to every action $\pi$ can take ($\mu(\frac{a}{\tau}) > 0$ whenever $\pi(\frac{a}{\tau}) > 0$).

\begin{table}[t]

\centering
\small
\setlength{\tabcolsep}{4pt}
\begin{tabular}{lcc}

\toprule
Method & Backbone & Exploration signal \\
\midrule
MAVEN        & VD & mutual information          \\
PAC        & MAPG & entropy    \\
FOX        & VD   & mutual information  \\
CMAE       & VD   & goal / entropy \\
EMC        & VD   & curiosity  \\
SUNRISE    & SAC  & variance \& Q-learning   \\
EMAX    & VD  & ensemble-voting   \\
\textbf{Ensemble MIX} & MAPG & \textbf{kurtosis \& actor logits}  \\
\bottomrule
\end{tabular}
\caption{Positioning of Ensemble MIX versus existing baselines for exploration in single-agent and MARL.}
\label{table:diff}

\end{table}

\subsubsection{Kurtosis.}
The kurtosis of a random variable is its standardized fourth moment \cite{westfall2014kurtosis}. For an ensemble
$\{Q^{\phi_{i,j}}(\tau_i,a_i)\}_{j=1}^N$ with sample mean $\bar Q$ the kurtosis is:
\begin{equation}
\kappa(\{Q^{\phi_{i,j}}\}) = \frac{\tfrac1N\sum_{j=1}^N(Q^{\phi_{i,j}}-\bar Q)^4}
{\big[\tfrac1N\sum_{j=1}^N(Q^{\phi_{i,j}}-\bar Q)^2\big]^2},
\end{equation}
and the excess kurtosis is $\kE(\tau_i,a_i)=\kappa(\{Q^{\phi_{i,j}}(\tau_i,a_i)\})-3$,
measured relative to the Gaussian reference value of $3$. Previous works from other domains made a similar use of the normal distribution to estimate the excess kurtosis. For instance, \cite{gabrielsen2015forecasting} estimated value at risk (VaR) in financial time series using excess kurtosis, measuring it relatively to the normal distribution. \cite{loperfido2020kurtosis} applied kurtosis for outlier detection in financial time series and used excess kurtosis as a criterion for portfolio selection.

\textbf{Epistemic and aleatoric uncertainty in MARL. } In MARL, aleatoric uncertainty, or inherent randomness \cite{lockwood2022review}, arises from various sources, including the stochastic dynamics of the environment, inherent noise, agents’ stochastic policies (e.g., epsilon-greedy), and policy shifts that introduce non-stationary \cite{lee2024pausing}. Epistemic uncertainty (lack of knowledge \cite{hofer2002approximate}), stems from limited familiarity with the environment’s model (e.g., rewards and transitions) or an incomplete understanding of agents’ policies in relation to one another. %MARL environments exhibit both aleatoric and epistemic uncertainty, arising from agents' stochastic exploration and their incomplete knowledge of other agents and the environment.

\textbf{Fisher bias correction.}  
The sample kurtosis computed from a finite ensemble of size $N$ is a biased estimator of the true excess kurtosis, and this bias becomes more severe as $N$ decreases. To correct for this effect, the Fisher adjustment \cite{joanes1998comparing} rescales the sample excess kurtosis to obtain an unbiased estimator under Gaussian assumptions  ($N \ge 4$, $\kE$ is the sample excess kurtosis):
\begin{equation}
\kappa_c(\tau_i, a_i)
=
\frac{N-1}{(N-2)(N-3)}
\Big[(N+1)\,\kE(\tau_i, a_i) + 6\Big].
\label{eq:fisher}
\end{equation}

\section{Related Work}
Ensemble MIX relates to three lines of research: MAPG with value decomposition, exploration in
MARL, and ensemble-based uncertainty estimation.

\textbf{MAPG and value decomposition.}
DOP~\citep{wang2020dop} and PAC~\citep{zhou2022pac} both combine a decomposed critic with decentralized actors. DOP employs a linear critic factorization to enable tractable off-policy learning, using off-policy Tree-Backup together with on-policy TD($\lambda$) updates. PAC extends the factorization with counterfactual predictions that provide additional assistive information and relies on maximum-entropy regularization to drive exploration. Ensemble MIX builds on the DOP architecture but differs from PAC by replacing entropy-based exploration with a selective uncertainty-driven exploration mechanism. In addition, unlike DOP, it replaces Tree-Backup with Retrace for off-policy correction.

\textbf{Exploration in MARL.}
Prior work addresses the challenge of large joint action space in several directions. \emph{Influence-based} methods encourage agents to visit states where their actions affect teammates: EITI and EDTI maximize mutual information and interaction value signals \citep{lee2024pausing}, while FOX promotes novel team formations through mutual-information-based rewards \citep{jo2024fox}. \emph{Goal-directed} methods explicitly guide exploration toward shared objectives; CMAE selects high-entropy goals from projected state spaces and trains agents to reach them \citep{pmlr-v139-liu21j}. \emph{Curiosity- and novelty-based} methods use intrinsic rewards or replay prioritization: EMC derives exploration bonuses from critic prediction errors and stores informative trajectories in episodic memory \citep{zheng2021episodic}. \emph{Committed and entropy-based} methods inject structured stochasticity into the policy: MAVEN conditions agents on a shared latent variable to induce temporally consistent exploration \citep{mahajan2019maven}, PAC and ADER drive exploration through shared and state-adaptive entropy regularization, respectively \citep{zhou2022pac, kim2023adaptive}. In contrast, we perform \emph{selective exploration}, restricting exploration to states and actions identified as uncertain by an ensemble-kurtosis signal. Table \ref{table:diff} summarizes the differences.

\textbf{Ensemble-based learning.}
Ensembles improve exploration and sample efficiency in both single- and multi-agent RL. EMAX \cite{learningensemble} combines ensembles with QMIX and value decomposition and uses voting for action selection. UCB \cite{chen2017ucb} uses ensemble variance as an exploration bonus in single-agent RL, and SUNRISE \cite{lee2021sunrise} extends this with combined exploration and uncertainty weighting.  Ensemble MIX differs from these approaches in two key aspects. First, it drives exploration via novel modification to the actor logits, whereas the aforementioned methods primarily inject exploration through the $Q$-function. Second, to address the identified problem of noise in MARL, we introduce ensemble kurtosis as a novel uncertainty signal, in contrast to prior work that relies mainly on established measures such as variance or voting-based aggregation.
\begin{figure}[t]
\centering
\begin{minipage}{0.46\columnwidth}
  \centering
  \includegraphics[width=\columnwidth]{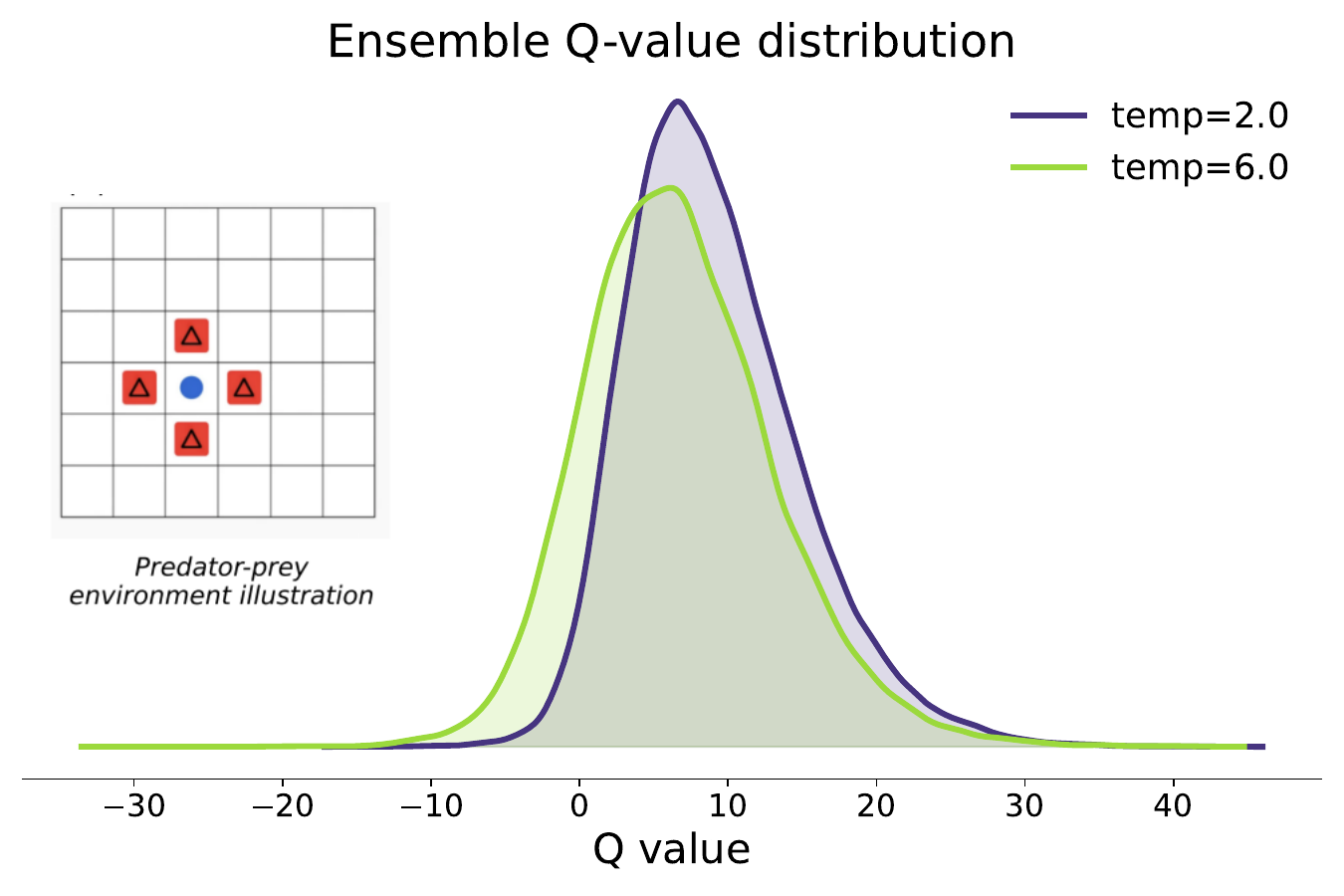}
  \captionof{figure}{Q-value distribution (KDE).}
  \label{fig:q-kde}
\end{minipage}
\hfill
\begin{minipage}{0.5\columnwidth}
  \centering
  \small
  \setlength{\tabcolsep}{4pt}
  \begin{tabular}{lcc}
  \toprule
  & Variance & Kurtosis \\
  \midrule
  $\tau=2$ & $(9.21 \pm 0.53) \times 10^{-4}$ & $2.525 \pm 0.024$ \\
  $\tau=6$ & $(13.27 \pm 1.15) \times 10^{-4}$ & $2.623 \pm 0.022$ \\
  $\%$ change & $44.0\%$ & $3.9\%$ \\
  \bottomrule
  \end{tabular}
  \captionof{table}{Aleatoric noise increase via temperature $\tau$ in predator-prey.}
  \label{tab:aleatoric-sweep}
\end{minipage}
\end{figure}
\section{The Ensemble MIX Algorithm}
This section presents our approach to enhancing sample efficiency in MARL settings via selective exploration. The key mechanism behind Ensemble MIX is: (i) an extension of standard MAPG with a decentralized ensemble of critics to produce an exploration signal, and (ii) an exploration component via a kurtosis-based signal that enables each agent to prioritize high-uncertainty regions during exploration through manipulation of the actor logits. We begin by presenting the architecture and then introduce the exploration mechanism. Pseudo-code is given in algorithm \ref{alg:main}.

\subsection{Ensemble of Critics}
Each agent's critic is an ensemble of $N$ sub-critics (Fig. \ref{fig:architecture}, component 1):
\begin{equation}
Q_i^{\phi_i}(\bm{\tau},a_i)=\frac1N\sum_{j=1}^N Q^{\phi_{i,j}}(\tau_i,a_i),
\end{equation}
and the global value is decomposed linearly~\citep{wang2020dop}:
\begin{equation}
\Qtot(\bm{\tau},\bm{a})=\sum_{i=1}^{K}\lambda_i(\bm{\tau})\,Q_i^{\phi_i}(\tau_i,a_i)+b(\bm{\tau}),
\end{equation}
with non-negative mixing weights $\lambda_i$ (the QMIX monotonicity constraint) and a
learnable bias $b$. Following standard practice in ensemble RL, diversity among members
is obtained through random initialization and distinct training mini-batches \cite{lee2021sunrise}. %All agents
%share the same ensemble, so the additional memory cost is $\mathcal{O}(NP)$ where each critic member in the ensemble has $P$ parameters (Table~\ref{tab:cost}).

\subsection{Exploration via Actor Logits and Kurtosis Signal}
We propose two variants of the exploration algorithm (Fig \ref{fig:architecture}, component 2). The first, and more selective, performs exploration only in regions of positive excess kurtosis, whereas the second performs continuous exploration via raw kurtosis. Nevertheless, both variants remain selective, as kurtosis guides exploration according to the tail shape of the ensemble distribution, assigning lower priority to light-tailed regions.

\textbf{Excess kurtosis variant. } Our algorithm applies kurtosis in two ways. First, each agent detects high-uncertainty states based on the mean kurtosis across all actions.  Second, once a relevant state is detected, prioritization is applied by weighting actions according to their kurtosis. At each iteration, the $i$-th agent computes the kurtosis over its $M_i$ actions:
\begin{equation}
\bar g_i\big(\tau_i,\{Q^{\phi_{i,j}}\}_{j=1}^N\big)=\frac{1}{M_i}\sum_{z=1}^{M_i}\kE(\tau_i,a_z).
\label{eq:gbar}
\end{equation}
If $\bar g_i>0$, the state is treated as high-disagreement and the agent perturbs its
logits by the per-action kurtosis; otherwise it uses the unperturbed policy. Formally,
given logits $\bar z^i=\{z_1^i,\dots,z_{M_i}^i\}$:
\begin{equation}
\tilde z_j^i=\begin{cases}
z_j^i+\beta\,\kE(\tau_i,a_j), & \text{if }\bar g_i>0,\\[2pt]
z_j^i, & \text{otherwise},
\end{cases}
\label{eq:explore}
\end{equation}
where $\beta$ is a hyperparameter, followed by a softmax. The condition $\bar g_i>0$
ensures that perturbation occurs only where the excess kurtosis exceeds the Gaussian
reference, that is, where the ensemble exhibits genuinely heavy-tailed disagreement. The
resulting per-state distribution, a softmax over $\tilde z^i$ followed by
$\epsilon$-greedy selection, is the policy that actually selects actions. We record it
as $\mu_i(\cdot|\tau_i)$. 

\textbf{Raw kurtosis variant. } In this variant, we substitute the excess kurtosis $\kE$ from Eq.~\ref{eq:explore} with the raw kurtosis $\kappa(\{Q^{\phi_{i,j}}(\tau_i, a_j)\})$. Since raw kurtosis is strictly positive for any distribution, the condition $\bar{g}_i > 0$ is always satisfied, and the logits are perturbed unconditionally at every step:
\begin{equation}
    \tilde{z}^i_j = z^i_j + \beta\, \kappa\!\left(\{Q^{\phi_{i,j}}(\tau_i, a_j)\}\right), \quad \forall\, j.
\end{equation}
Despite being less selective than excess kurtosis, the raw variant remains focused on heavy-tailed deviations and assigns lower probability to disagreement concentrated around the mean Q-value. This yields a signal that performs continuous exploration yet remains more selective than techniques that explore the entire state space with similar probability.

\textbf{Quantifying changes in behavior policy. } Since the kurtosis perturbation directly alters the behavior
policy used to collect data, we quantify how much any
single action's probability can shift as a result. The following
lemma bounds this change in terms of the perturbation strength.  Let $p_0(a_j) := \mathrm{softmax}(z^i)_j$ and
$p(a_j) := \mathrm{softmax}(\tilde z^i)_j$ denote the original and
perturbed probability of action $a_j$ at state $\tau_i$, and let the perturbation spread at $\tau_i$ be:
\begin{equation}
    \Delta(\tau_i) \;:=\; \beta\Big(\max_j \kappa_j - \min_k \kappa_k\Big).  
\end{equation}
\begin{figure*}[t]
\centering
\includegraphics[width=0.99\textwidth]{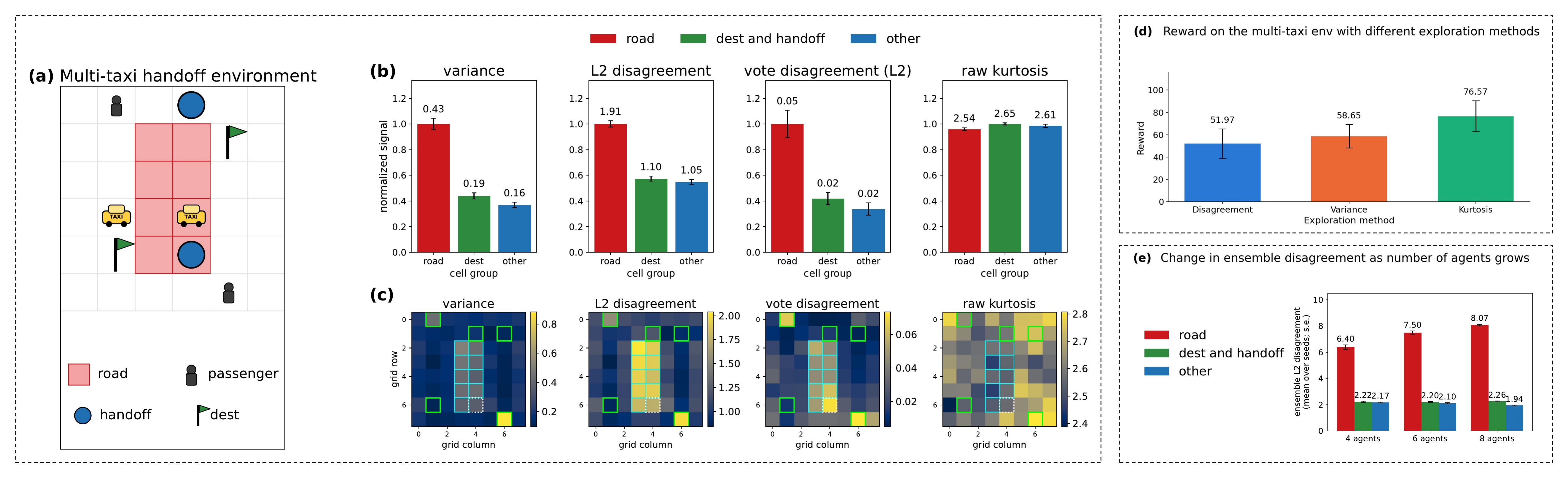}

    \caption{\textbf{Analysis of aleatoric noise and ensembles on the multi-taxi environment.}
    \textbf{(a)} $8\times8$ multi-taxi environment illustration.
    \textbf{(b-c)} Common ensemble measurements on a heatmap that represent values on the grid. \textbf{(d)} The reward thorough training with different exploration methods
    \textbf{(e)} Increasing the number of agents with effect on ensemble disagreement. }
\label{fig:toy1}
\end{figure*}
\begin{lemma}
\label{lem:max-prob-ratio}
For every action $a_j$,
\[
  e^{-\Delta(\tau_i)} \;\le\; \frac{p(a_j)}{p_0(a_j)} \;\le\; e^{\Delta(\tau_i)}.
\]
\end{lemma}
The perturbed probability of any action differs from its
original value by at most a multiplicative factor of
$e^{\Delta(\tau_i)}$, tying the behavior policy's deviation directly
to the kurtosis spread at that state.
The same lemma holds for both the raw and excess kurtosis variants.
\subsection{Retrace-Based Critic and IS Correction}
We retain DOP's on-/off-policy critic stucture (Eq.~\ref{eq:dop_loss}) and replace only the
off-policy target, which becomes:
\begin{align}
y^{\text{off}}=&\,\Qtotp(\bm{\tau},\bm{a})
+\sum_{t=0}^{m-1}(\gamma\lambda)^t C_{\bm{\tau}}
\big[r_t \nonumber\\
&+\gamma\E_{\bm{a}_{t+1}\sim\bm{\pi}}[\Qtotp(\bm{\tau}_{t+1},\bm{a}_{t+1})]
-\Qtotp(\bm{\tau}_t,\bm{a}_t)\big],
\label{eq:semus_off}
\end{align}
where, in place of DOP's TB  coefficient: $\lambda\bm{\pi}(\bm{a}_l|\bm{\tau}_l)$, we use the Retrace coefficient (Fig \ref{fig:architecture}, component 3):
\begin{equation}
C_{\bm{\tau}}=\prod_{l=0}^{t}\min\!\Big(1,\ \frac{\bm{\pi}(\bm{a}_l|\bm{\tau}_l)}{\bm{\mu}(\bm{a}_l|\bm{\tau}_l)}\Big),
\qquad \bm{\mu}=\prod_{i=1}^{k}\mu_i,
\label{eq:retrace}
\end{equation}
with $\mathcal{L}^{\text{off}}(\phi)=\E_{\bm{\mu}}[(y^{\text{off}}-\Qtot(\bm{\tau},\bm{a}))^2]$
and $\mathcal{L}^{\text{on}}$ as in DOP.

\textbf{The role of the behavior policy.}
The IS correction accounts for the difference between the behavior policy
$\bm{\mu}$ that collected the data and the actor $\bm{\pi}$ that we evaluate. In our
setting, $\bm{\mu}$ and $\bm{\pi}$ differ, since the kurtosis perturbation shifts action
away from $\bm{\pi}$, which requires the collected data off-policy to be corrected. $\bm{\mu}=\prod_i\sigma(\tilde z^i)$ is the kurtosis-perturbed,
policy of Eq.~\ref{eq:explore}, used as the denominator in Eq.~\ref{eq:retrace}.
 \begin{table*}
\centering
\begin{tabular}{lccccc}
\toprule
 & \makecell{Ensemble MIX,\\ smaller\\ ensemble ($N=5$)} & \makecell{Ensemble MIX,\\ without\\ retrace} & \makecell{No exploration\\ ($\beta = 0$)} & \makecell{Ensemble MIX\\ with variance\\ exploration} & \makecell{Original, all\\ components\\ (with exploration)} \\
\midrule
1c3s5z & $0.63 \pm 0.06$ & $0.79 \pm 0.04$ & $0.88 \pm 0.02$ & $0.82 \pm 0.01$ & $0.96 \pm 0.00$ \\
MMM2   & $0.67 \pm 0.08$ & $0.77 \pm 0.06$ & $0.83 \pm 0.05$ & $0.8 \pm 0.04$  & $0.94 \pm 0.04$ \\
MMM3   & $0.68 \pm 0.04$ & $0.69 \pm 0.03$ & $0.75 \pm 0.05$ & $0.69 \pm 0.03$ & $0.82 \pm 0.03$ \\
MACF   & $0.64 \pm 0.04$ & $0.66 \pm 0.03$ & $0.72 \pm 0.05$ & $0.71 \pm 0.03$ & $0.82 \pm 0.03$ \\
\bottomrule
\end{tabular}
\caption{Ablation and analysis. The leftmost and second columns show the performances with a smaller ensemble. In the third column, we add the ensemble with no exploration. The fourth is an ensemble mix with variance-based exploration.}
\label{table:ablation}
\end{table*}
\section{Algorithm Motivation and Analysis}
This section presents the three main research questions that motivate our work:
%\subsection{Algorithm Motivation}

\textbf{1. What advantages do ensemble methods provide for learning and exploration in MARL over existing techniques?}   Standard exploration methods from single-agent RL, such as entropy maximization \cite{haarnoja2018soft}, often scale poorly to MARL. By promoting random exploration across a large joint action space, they often lead agents toward redundant regions of the state space and can ultimately hinder performance \cite{kim2023adaptive, jo2024fox}. Ensembles, on the other hand, are considered highly effective at identifying regions of high epistemic uncertainty. When used correctly, they can guide exploration toward underexplored regions and allow the learning algorithm to avoid redundant exploration \cite{zisselman2023explore, liu2024ovd}. They are also known to exhibit strong sample efficiency, which is essential for MARL settings.
\begin{figure*}[t]
\centering
\includegraphics[width=0.99\textwidth]{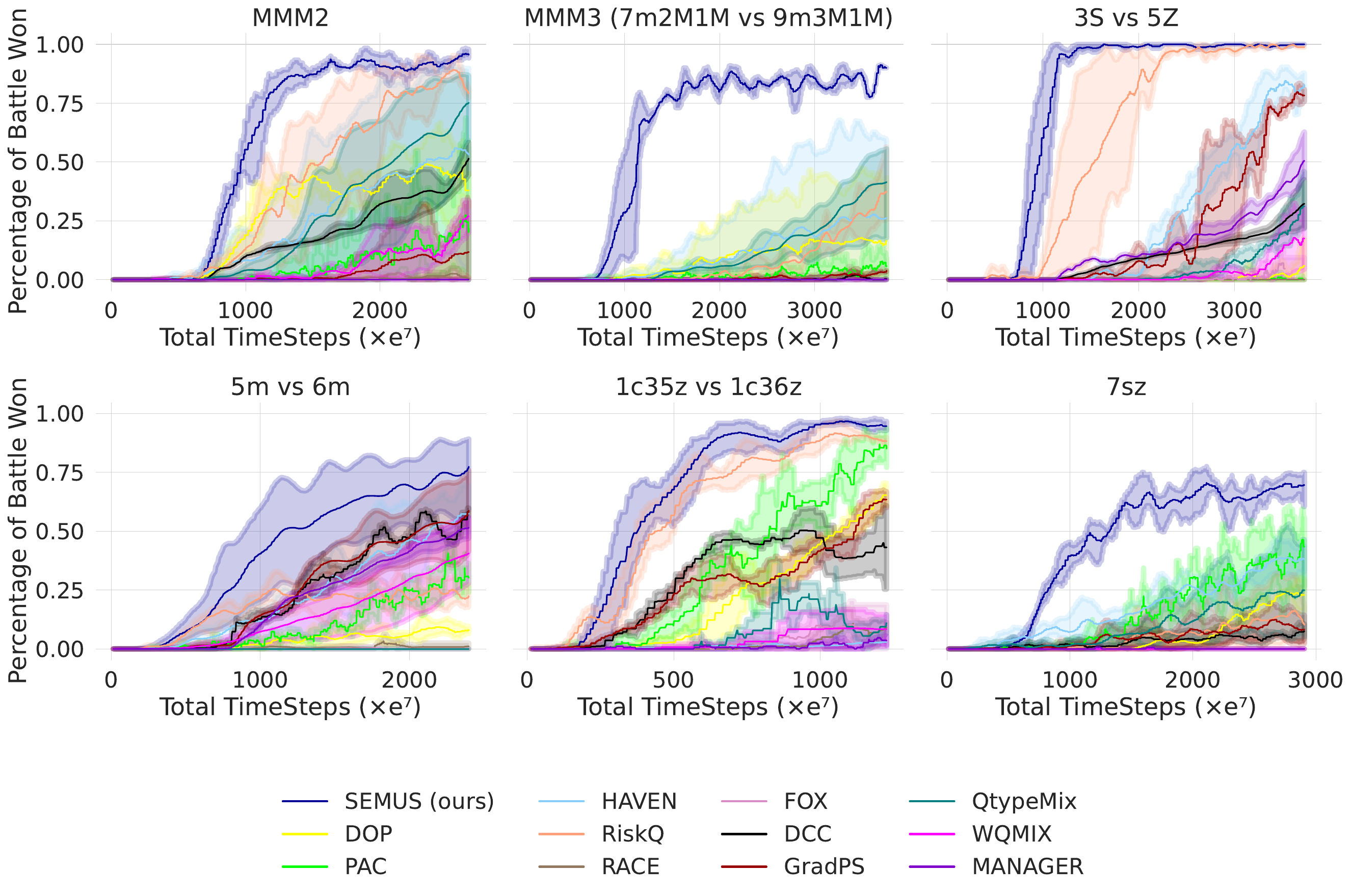}
\caption{\textbf{(a):} Win rates on six (regular-reward) SMAC maps. \textbf{(b):} Win rates on six sparse reward maps. Sparse reward structure is similar to previous works and is given in the Appendix C.}
\label{fig:smac}
\end{figure*}

\textbf{2. Does noise become more dominant in ensemble predictions in MARL?} To illustrate this problem,  we examine the multi-taxi environment (Fig. \ref{fig:toy1}a). The environment is constructed from a $8\times 8$ grid-world in which two taxis cooperatively deliver a passenger with several pairs of taxis and an infinite supply of passengers. Each taxi is constructed from an actor and a tabular ensemble of Q-functions, initialized by sampling a random value from a normal distribution. In each episode, a passenger is carried by the first taxi to a handoff location and then picked up by a second taxi to a final destination. Both taxis must cross a central region (red) that mimics a busy road by imposing a random penalty $-\,\mathrm{Laplace}(0, 8)$ when occupied by two or more taxis, while successful delivery and handover yield a reward $+10$. This introduces aleatoric uncertainty that stems from the stochastic policies of co-evolving exploring agents. The handoff and destination cells represent epistemic uncertainty. Figs. \ref{fig:toy1}b-c show that common signals, including ensemble variance, L2 disagreement, and voting, become increasingly dominated by noisy regions, driving the ensemble measurements toward irreducible uncertainty. Green rectangles on the heat map represent handoff, destination, and busy road cells. Fig.~\ref{fig:toy1}d shows how this negatively affects reward thorough training when exploration signals such as variance and disagreement are added to the actor logits. In Fig.~\ref{fig:toy1}e, we increase the number of agents in the environment, introducing more co-evolving taxis that step through the road,  which can be seen to amplify noise and ensemble disagreement on those areas. In Appendix E, we show this phenomena persists across different noise distributions and additional grid environments.

\textbf{3. Is Kurtosis a suitable signal for MARL settings? }
Following the previous two motivational questions,
it can be concluded that for ensemble-based methods to enable effective exploration in MARL, the exploration signal should: (i) Account for the large joint action space and selectively focus only on a subset of states and actions, and (ii) mitigate the impact of aleatoric noise. We argue that kurtosis is well-suited to this role. First,  as a measure of distribution tailness \cite{westfall2014kurtosis}, kurtosis emphasizes large deviations from the ensemble mean Q-value, which captures regions of strong disagreement and uncertainty. Our empirical analysis on the multi-taxi  shows that this strong disagreement is observed in the areas of epistemic uncertainty. To asses how well kurtosis cope with high noise, we analyze its sensitivity to such settings in Figs. \ref{fig:toy1}b-d (over the area of the noisy road), which show a mild effect on its predictions when noise and aleatoric uncertainty are increased, with higher reward than baselines. \newline
\textbf{Non-tabular settings.} To show these problems persist beyond the tabular settings, we experiment with the Predator-Prey (PP) environment \cite{zhou2022pac} where agents collaborate to catch prey by surrounding it. The agents are trained with our proposed ensemble-based exploration algorithm. We progressively increase the softmax temperature of all actors to induce higher stochasticity in action selection (Table \ref{tab:aleatoric-sweep}) and noisier collaboration. The corresponding ensemble distributions throughout training are illustrated in the accompanying figure. It can be seen that higher $\tau$ leads to wider spread but with a mild effect on the distribution tail, as recorded in kurtosis values. This explains why kurtosis remains less affected by higher randomness, with lower susceptibility to over-exploration as $\tau$ increases.

%\textbf{Notation. } We model the critic estimates at a fixed
%$(\tau_i,a_i)$ as:
%\begin{equation}
%Q^{\phi_{i,j}}=\mu_j+\epsilon_j+\delta_j,\qquad j=1,\dots,N,
%\label{eq:mixmodel}
%\end{equation}
%where $\epsilon_j\sim F(0,\sigma_j^2)$ is \emph{aleatoric} noise with mean $0$, variance
%$\sigma_j^2$, and excess kurtosis $\kappa_0$, and $\delta_j\in\{0,\Delta\}$ is an
%\emph{epistemic} shift with $P(\delta_j=\Delta)=p_j$ and $\bar p=\tfrac1N\sum_j p_j$.
%Here $\sigma_j^2$ is irreducible noise, each critic fits a different noise
%realization, while $(\Delta,\bar p)$ encode reducible disagreement.

\section{Evaluation and Discussion}
This section assesses the performance of Ensemble MIX across a range of settings: 13 SMAC maps ~\cite{samvelyan2019starcraft}), two predator-prey grids~\cite{zhou2022pac}, and multi-agent car following
(MACF)~\cite{shen2023riskq}.
Hyperparameters and network
architectures are fixed across experiments (Appendix~D). Each run uses 7 seeds with
$95\%$ confidence intervals. We compare Ensemble MIX against $11$ baselines from different categories. 
\textbf{MAPG:} DOP~\cite{wang2020dop}, PAC~\cite{zhou2022pac} . \textbf{Efficient learning:}  GradPS~\cite{qin2025gradps}, MANAGER~\cite{chen2025novelty}. \textbf{Exploration:} FOX~\cite{jo2024fox}. \textbf{Classical:} Weighted
QMIX~\cite{rashid2020weighted}. \textbf{Uncertainty-aware:} RiskQ~\cite{shen2023riskq}. \textbf{Hierarchical learning:} HAVEN~\cite{xu2023haven}, DCC~\cite{li2025coordinating}. \textbf{General:} QtypeMix~\cite{fu2025qtypemix}, RACE~\cite{li2023race}. EMAX omitted due to unreleased code; all other baselines use authors' official code. \newline
\textbf{Implementation details. } The ensemble size is
$N=10$, and each sub-critic is a two-layer fully connected network. All experiments are with the raw kurtosis variant, comparison to excess kurtosis results is given in the Appendix. We use standard ensemble diversity techniques based on random initialization and different sample batches.  Full details are given in Appendix D. All source code will be made available on publication.

\textbf{Q1: How does Ensemble MIX perform on standard MARL settings?}
To answer this question, we conduct experiments on six SMAC maps, presented in Fig.~\ref{fig:smac}a. Ensemble MIX consistently outperforms the baselines, with the
largest margins on the super-hard MMM2 and MMM3 maps, which require diverse per-agent
strategies with different types of agents. RiskQ achieves the second best performance on four out of the six maps, with QtypeMix third best performances. 
\begin{figure*}[t]
\centering
\includegraphics[scale=0.24]{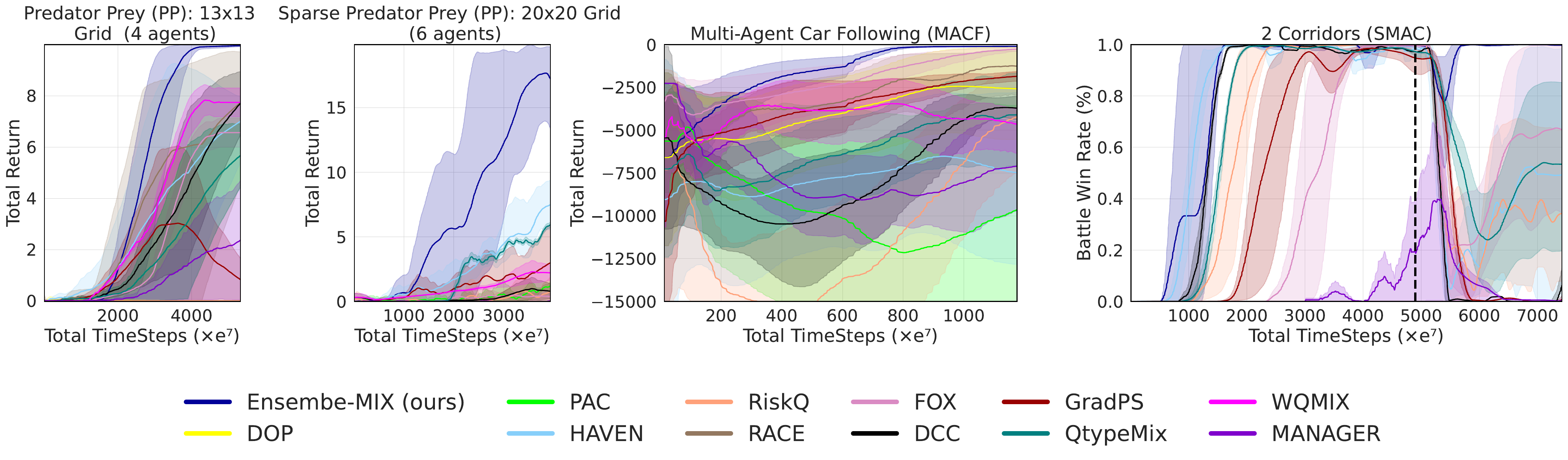}
\caption{Returns and win rates on predator-prey, MACF, and the two-corridors (SMAC) map.}
\label{fig:other}
\end{figure*}

\textbf{Q2: How does Ensemble MIX perform on exploration-intensive settings?} To answer this question, we experiment with: (i) Six sparse-reward SMAC maps (Fig.~\ref{fig:smac}b), we use similar sparse reward structure as prior works ~\cite{jo2024fox, liu2024imagine}, which is detailed in Appendix C. (ii) The two-corridor map, where 2 allies fight an enemy unit, with the short corridor closed midway through training (doted line in 2 corridor graph in Fig. \ref{fig:other}), forcing
adaptation to changing terrain. Ensemble MIX fully adapts after $0.1$M, faster than all baselines methods. (iii) Sparse predator-prey (PP), where four agents (instead of 2) must coordinate on every
capture. As the grid size increases, the reward becomes sparser, and only Ensemble MIX solves the task.

\textbf{Q3: How does Ensemble MIX perform on noisy environments?} To answer this question, we experiment with the Car-following environment (MACF). In MACF, two agents drive toward a goal while remaining within sight of each other.
Over-speeding causes a crash with some probability, introducing aleatoric
noise in over-speed actions. Ensemble MIX attains a higher reward (Fig.~\ref{fig:other}), avoiding over-speed events. %This is the empirical counterpart
%of Theorem~\ref{thm:main}: a variance signal would over-explore the noisy high-speed
%regime, whereas kurtosis attenuates it, so Ensemble MIX does not pursue irreducible noise.

\textbf{Ablations and parameter analysis.}
Table ~\ref{table:ablation}  reports the cumulative contribution of each component by progressively increasing ensemble size, and adding kurtosis-based exploration. To verify the effectiveness of kurtosis, we additionally substitute kurtosis with variance for exploration while leaving the remaining architecture unchanged. %Table \ref{tab:beta_sweep} in the appendix examines the effect of different values of $\beta$, showing that excessive exploration can also degrade performance.

%\textbf{Bias correction experiments.} Sample kurtosis can be biased for small ensemble sizes. We therefore evaluate an ablation that applies the Fisher correction (Eq.~4) to the exploration signal, used only in this experiment and not in the main results. Table~\ref{tab:bias_correction} compares both versions, showing that the correction helps only in some cases.

\textbf{Limitations (computational resources and ensemble overhead). }  As all critics share the same ensemble (standard approach in CTDE), assuming each critic network has $P$ parameters, the memory overhead is is $\mathcal{O}(NP)$ for the entire system. %Table~\ref{tab:cost} (appendix D) shows memory and latency comparison on our GPU compared to ensemble of size $1$.  

%\textbf{Additional results and empirical validation. } Appendix E (Figs. \ref{fig:toy_grid},\ref{fig:toy_grid_normal}, and \ref{fig:relay8-noise}) provides additional experiments on multi-taxi and other grid world setting. We show that exploration via kurtosis guides the agents to areas of epistemic uncertainty, while largely avoiding over-exploration of noisy areas. 

%\textbf{Formal results.} Theorem~\ref{thm:main} shows the excess kurtosis converges to a value
%depending only on the disagreement probability $p$ (the fraction of members
%carrying the epistemic shift), namely $\bigl(1-6p(1-p)\bigr)/\bigl(p(1-p)\bigr)$.
%This is finite for any $p \in (0,1)$ and grows large only as $\Delta \to \infty$
%with $p \to 0$ or $p \to 1$, where almost all members agree while one carries an
%unboundedly large shift.
%This rarely arises, since shifts are finite in practice and random
%initialization with distinct minibatches keeps $p$ away from these extremes.

%\textbf{Generalizing Theorem \ref{thm:main}. } Theorem~\ref{thm:main} extends naturally to the case where each ensemble member is associated with its own epistemic shift $\Delta_j$, drawn i.i.d. and independently of the disagreement pattern. The proof proceeds unchanged after substituting $\Delta^2$ and $\Delta^4$ with $\mathbb{E}[\Delta^2]$ and $\mathbb{E}[\Delta^4]$.

\section{Conclusion}
We propose Ensemble MIX, a sample-efficient MARL algorithm that performs selective exploration in large action-state space while accounting for the high aleatoric noise present in MARL. To achieve this, we propose a novel algorithm for extending MAPG with ensemble of critics, and directly manipulating the actor logits using an ensemble-derived exploration signal. Our work identifies a major challenge in applying ensemble techniques to MARL: noise and aleatoric uncertainty dominate ensemble predictions due to the presence of multiple exploring agents. We study this effect on common ensemble measurements such as voting and variance, and propose a novel technique to quantify ensemble uncertainty via kurtosis, which, as our analysis show, exhibit less sensitivity to noise in high-uncertainty settings.  Ensemble MIX achieves consistent improvements over state-of-the-art baselines across exploration-intensive and sparse reward SMAC maps, as well as on general MARL settings.

\textbf{Future work. } Future work could extend the kurtosis-driven exploration beyond the DOP backbone to value-based and other MAPG methods with continuous-action space. Another direction is adaptively tune the perturbation strength $\beta$ via learned parameters rather than fixing it.
%Bibliography
\bibliographystyle{unsrt}  
\bibliography{references}

\newpage
\appendix

\section{Algorithm\label{appendix:algo}}

Action selection (Algorithm~\ref{alg:select}) perturbs the logits only when the mean
excess kurtosis is positive; the resulting distribution is logged as $\bm{\mu}$ and reused
by the Retrace critic. The main loop (Algorithm~\ref{alg:main}) trains the critic with
Eq.~\ref{eq:dop_loss}, using the Retrace coefficient of Eq.~\ref{eq:retrace} in the
off-policy term, and updates the actors with the mixed on-/off-policy
gradient~\citep{gu2017interpolated}. The on-policy minibatch is smaller than the
off-policy one ($|N_1|<|N_2|$).

\begin{algorithm}[h]
\caption{Ensemble MIX}
\label{alg:main}
\begin{algorithmic}
\STATE \textbf{Input:} episodes $L$, target interval $X$, episode length $T$, agents $K$
\STATE Initialize mixing network $\Qtot$, per-agent ensembles $\{Q^{\phi_{i,j}}\}_{j=1}^N$, actors $\{\pi_{\theta_i}\}$
\STATE Initialize targets $\Qtotp,Q^{\phi'_{i,j}}$ and buffers $\mathcal{D}_{\text{on}},\mathcal{D}_{\text{off}}$
\FOR{episode $=1$ to $L$}
  \FOR{$t=1$ to $T$}
    \STATE Obtain $\bm{a}_t$ and behavior probabilities $\bm{\mu}(\bm{a}_t|\bm{\tau}_t)$ via Alg.~\ref{alg:select}
    \STATE Execute $\bm{a}_t$; observe $r_t,s_{t+1}$; store $(s_t,\bm{a}_t,r_t,s_{t+1},\bm{\mu})$
    \STATE Sample $N_1\sim\mathcal{D}_{\text{on}}$, $N_2\sim\mathcal{D}_{\text{off}}$
    \STATE Update critics with Eq.~\ref{eq:dop_loss} (Retrace coeff., Eq.~\ref{eq:retrace})
    \STATE Update actors with the mixed on-/off-policy gradient
    \IF{$t\bmod X=0$}
      \STATE Update target networks
    \ENDIF
  \ENDFOR
\ENDFOR
\end{algorithmic}
\end{algorithm}

\begin{algorithm}[h]
\caption{Action Selection}
\label{alg:select}
\begin{algorithmic}
\STATE \textbf{Input:} actors $\{\pi_i\}$, ensembles $\{Q^{\phi_{i,j}}\}$, history $\bm{\tau}$, action counts $\{M_i\}$
\STATE \textbf{Output:} actions $\bm{a}$ and behavior probabilities $\bm{\mu}$
\FOR{agent $i=1$ to $K$}
  \STATE $\bar z^i=f_{\theta_i}(\tau_i)$; compute $\bar g_i$ (Eq.~\ref{eq:gbar})
  \STATE Perturb $\bar z^i$ via Eq.~\ref{eq:explore} if $\bar g_i>0$, obtaining $\tilde z^i$
  \STATE $\mu_i\leftarrow\epsilon\text{-greedy}(\sigma(\tilde z^i))$; sample $a_i\sim\mu_i$
\ENDFOR
\end{algorithmic}
\end{algorithm}

\section{Training Parameters and computation resources\label{appendix:parameters}}
Configurations and hyper-parameters values are provided in table \ref{tab:hyperparam}.
\begin{table*}[ht]
  \centering
  \caption{Hyperparameter Table - SMAC}
    \begin{tabular}{lr}
    \toprule
    \multicolumn{2}{l}{\textbf{ Algorithm Parameters}} 
    \\
    \midrule
    \textbf{HyperParameters} & \textbf{Value} \\
    $T$ (train interval) &  20K \\
    $U$ (num test episodes) &  24\\
    $C_1$ &  0.01  \\
    $C_2$ &  0.001 \\
    $\beta$ &  0.001 \\
    $v$ &  0.5 \\
    $N_1$ &  32 \\
    $N_2$ & 5000  \\ ensemble size (N) & 10\\  
    \midrule
    
    \bottomrule
    \end{tabular}%
  \label{tab:hyperparam}%
\end{table*}

All experiments are conducted on stations with 64 Intel(R) Xeon(R) CPU E5-2683 v4 @ 2.10GHz, 
251G RAM, and NVIDIA GeForce GTX 1080 Ti 12GB.
\section{Starcraft $\RN{2}$ scenarios \label{appendix:maps}}
Our main focus is on scenarios that require the agents to learn diverse skills or exhibit some level of exploration to solve the task. We use the term diversity to describe inter-agent diversification, meaning different agents acquiring different sets of skills. Note that then the state space in all SMAC maps is augmented with the unit type, which allows the agents to acquire different roles based on the unit type. \newline
\textbf{MMM2.} Two identical teams battle each other, aiming to eliminate the units of the opposing team. Each team controls $3$ types of units. The first type is a flying, healing unit called Medivac, capable of restoring health to nearby units. The second are the Marines units, which are capable of attacking both ground and air enemy units, and third, the Marauders, bulkier units only capable of attacking enemy ground units. The MMM2 scenario requires diversity among agents, where each type of agent needs to learn a different set of strategies. \newline
\textbf{MMM3 (7m2M1M vs 9m3M1M) .} This is a ultra hard version of the MMM2 map. The allies control a smaller team of 10 agents compared to the larger enemy team of 13 agents.\newline
\textbf{1c3s5z vs 1c3s6z.}A scenario where both teams control 3 types of agents, Colossus, Stalkers, and Zealots. The enemy team has a higher number of Zealots. \newline
\textbf{3s vs 5z. } In this scenario, 3 Stalkers (ranged units) are controlled by agents facing 5 Zealots (melee units). The challenge lies in overcoming the numerical disadvantage by leveraging Stalkers' kiting abilities, coordinating attacks, and maintaining effective positioning to avoid Zealots' close-range damage.

\section{Implementation \& Reproducibility}
All source code will be made available on publication under an open-source license. We refer the reader to the included README file, which contains instructions to recreate the experiments discussed in this paper. 

\section{Epistemic and aletoric uncertainty in multi-agent RL \label{APPENDIX:uncer}}
In MARL environments, aleatoric uncertainty, or inherent randomness \cite{lockwood2022review}, arises from various sources, including the stochastic dynamics of the environment, inherent noise, agents’ stochastic policies (e.g., epsilon-greedy), and policies shift that introduce non-stationarity. Epistemic uncertainty (lack of knowledge \cite{hofer2002approximate}), on the other hand, stems from limited familiarity with the environment’s model (e.g., rewards and transitions) or an incomplete understanding of agents’ policies in relation to one another. The presence of multiple agents within the same environment introduces both aleatoric and epistemic uncertainty, as agents’ exploration and stochastic policies induce aleatoric uncertainty \cite{lockwood2022review} and unfamiliarity of agents with other agents' models and the environment is epistemic. 

In our uncertainty-weighted critic, we apply the weighting by utilizing  
ensemble kurtosis, i.e., disagreement between ensemble members, which is usually associated with epistemic uncertainty. Yet, the noise in our settings stems both sources: randomness in agent policies (i.e., aleatory) and epistemic uncertainty originated from the unfamiliarity of agents with the model of one another. Similar concerns have been addressed by recent works \cite{lee2021sunrise}, which showed ensemble variability to be effective in capturing aleatoric uncertainty. \cite{lee2021sunrise} injected noise into the environment and detected noisy states via ensemble variability. Motivated by those results, we apply the ensemble approach in our settings, which mixes both epistemic and aleatoric sources.

\begin{figure}[t]
    \centering
    \includegraphics[width=0.77\columnwidth]{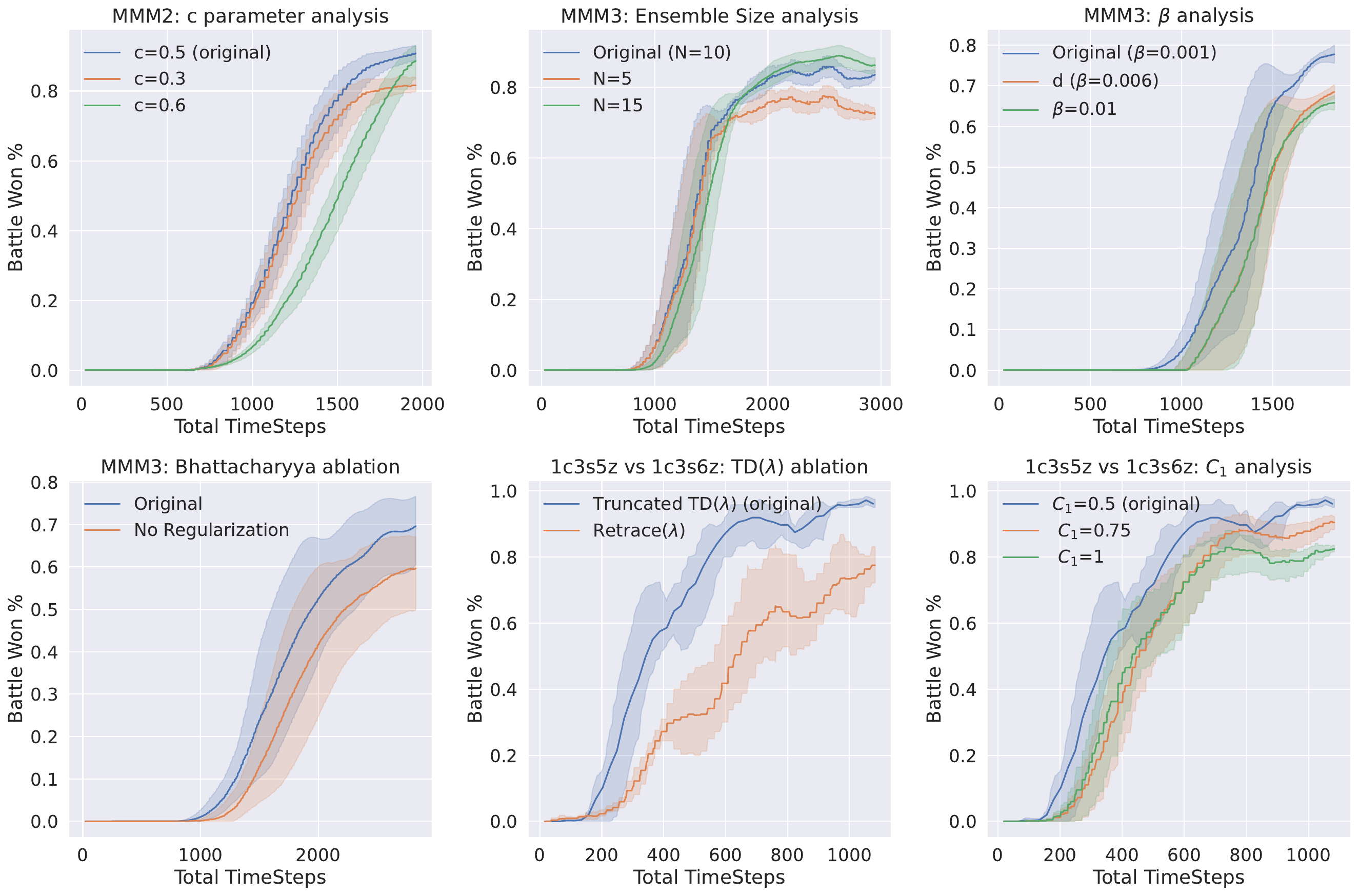}
    \caption{Parameter analysis on MMM2 and MMM3 maps for $\mathbf{c},\beta, C_1$, and ensemble sizes. Ablations on the Bhattacharyya distance and our version of TD($\lambda$) versus retrace($\lambda$) are presented in the bottom graphs.}
\label{fig:ablation3}
\end{figure}

\end{document}